\documentclass[5p,times]{elsarticle}

\usepackage{siunitx}
\usepackage{microtype}
\usepackage{amsmath} 
\usepackage{amssymb}  
\usepackage{graphicx}

\usepackage{amsthm}
\usepackage{booktabs}
\usepackage{algorithm} 
\usepackage{algpseudocode} 
\usepackage{enumitem}

\usepackage{cuted}
\usepackage{multirow}
\usepackage{eso-pic}

\newtheorem{procedure}{Procedure}

\newcommand{\real}[1]{\text{Re}\left[#1\right]}
\newcommand{\imag}[1]{\text{Im}\left[#1\right]}

\newcommand{\tx}{\text{Re}\left[t\right]}
\newcommand{\ty}{\text{Im}\left[t\right]}
\newcommand{\txk}{\text{Re}\left[t_k\right]}
\newcommand{\tyk}{\text{Im}\left[t_k\right]}

\newcommand{\vc}{\mathbf{c}}
\newcommand{\vcp}{\mathbf{c}^\prime}
\newcommand{\vt}{\mathbf{t}}

\newcommand{\figref}{Figure~}

\newcommand{\tabref}{Table~}

\AddToShipoutPictureBG*{%
    \AtPageUpperLeft{%
        \setlength\unitlength{1in}%
        \hspace*{\dimexpr0.5\paperwidth\relax}
        \makebox(0,-1.5)[c]{%
            \begin{tabular}{c}
                J.S. van Hulst \emph{et al.}, ``Tilt-based Aberration Estimation in Transmission Electron Microscopy.''\\
                Preprint (revised version). This manuscript is under peer review. Please cite the published version when available.
            \end{tabular}%
        }%
    }%
}%

\AddToShipoutPictureBG*{%
\AtPageUpperLeft{%
\setlength\unitlength{1in}%
\hspace*{\dimexpr0.5\paperwidth\relax}
\makebox(0,-22.3)[c]{
\footnotesize
\begin{tabular}{c c}
\copyright~The Authors
\end{tabular}}}}

\begin{document}
\begin{frontmatter}
    \title{Tilt-based Aberration Estimation in Transmission Electron Microscopy\tnoteref{t1}}
    \tnotetext[t1]{This work is part of the research project entitled \emph{Learning in Motion}, a collaboration between the Eindhoven University of Technology and several industry partners. This project is co-financed by Holland High Tech, top sector High-Tech Systems and Materials, with a PPP innovation subsidy for public-private partnerships for research and development.}

    \affiliation[tue]{organization={Department of Mechanical Engineering, Control Systems Technology Section},
            addressline={Eindhoven University of Technology}, 
            city=Eindhoven,
            address={PO Box 513, 5600MB,},
            country={the Netherlands}}
\affiliation[tfs]{organization={Thermo Fisher Scientific},
            city=Eindhoven,
            address={de Schakel 2, 5651GH,},
            country={the Netherlands}}
\author[tue]{Jilles S. van Hulst\corref{cor1}} 
\cortext[cor1]{Corresponding author. \emph{Email address:} {\tt j.s.v.hulst@tue.nl}}
\author[tfs]{Erik M. Franken}
\author[tfs]{Bart J. Janssen}
\author[tue]{W.P.M.H. (Maurice) Heemels}
\author[tue]{Duarte J. Antunes}


\begin{abstract}                
Transmission electron microscopes (TEMs) enable atomic-scale imaging but suffer from aberrations caused by lens imperfections and environmental conditions, reducing image quality. These aberrations can be compensated by adjusting electromagnetic lenses, but this requires accurate estimates of the aberration coefficients, which can drift over time.
This paper introduces a method for the estimation of aberrations in TEM by leveraging the relationship between an induced tilt of the electron beam and the resulting image shift. The method uses a Kalman filter (KF) to estimate the aberration coefficients from a sequence of image shifts, while accounting for the drift of the aberrations over time. The applied tilt sequence is optimized by minimizing the trace of the predicted error covariance in the KF, which corresponds to the A-optimality criterion in experimental design.
We show that this optimization can be performed offline, as the cost criterion is independent of the actual measurements. The resulting non-convex optimization problem is solved using a gradient-based, receding-horizon approach with multi-starts. Additionally, we develop an approach to estimate specimen-dependent noise properties using expectation maximization (EM), which are then used to tailor the tilt pattern optimization to the specific specimen being imaged.
The proposed method is validated on a real TEM set-up with several optimized tilt patterns. The results show that optimized patterns significantly outperform naive approaches and that the aberration and drift model accurately captures the underlying physical phenomena. A direct comparison with the widely used Zemlin tableau shows that the proposed method achieves comparable or higher image quality on amorphous specimens, while additionally extending to non-amorphous specimens where the Zemlin tableau cannot operate.
\end{abstract}


\end{frontmatter}


\section{Introduction}
Transmission electron microscopy (TEM) has enabled and continues to enable scientific progress across numerous fields, including materials science, semiconductors, and life sciences. A notable example of its impact is in the development of the COVID-19 vaccine, where TEM played a critical role by enabling the visualization of the virus' spike protein~\citep{Walls2020}.
What makes the TEM so powerful is its ability to produce images at the atomic scale. This capability stems from the use of electrons, which have significantly shorter wavelengths than visible light. Since resolution is inversely proportional to wavelength, the shorter the wavelength, the higher the attainable resolution.

\begin{figure}[!t]
    \centering
    \includegraphics[width=0.485\textwidth]{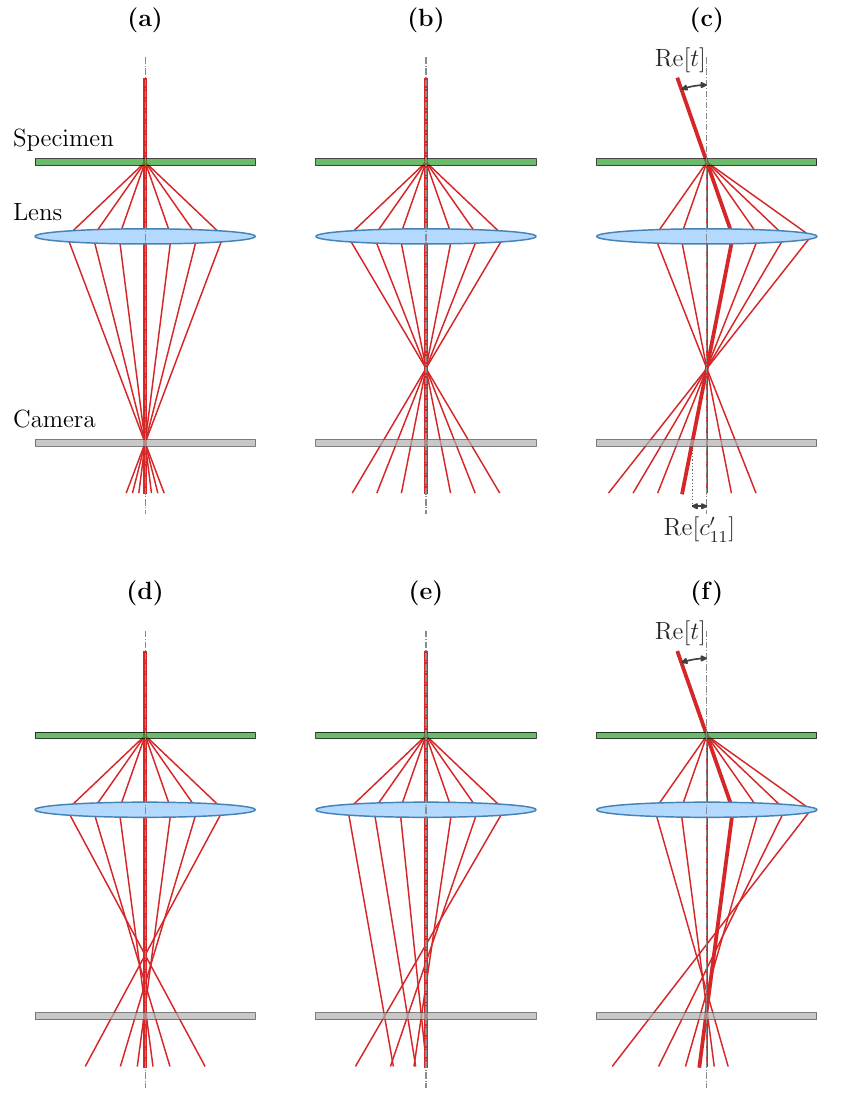}
    \caption{Schematic illustration of beam tilt and aberrations in TEM. In each panel, an electron beam (red) hits the specimen (green) at a point on the optical axis (dash-dotted line). The scattered electrons are represented by seven rays (red). These rays propagate through an electromagnetic lens (blue) and arrive at the camera (grey). \textbf{(a)}~Aberration-free case: all rays converge to a single point. \textbf{(b)}~Defocus: the rays converge before the camera plane. \textbf{(c)}~Beam tilt combined with defocus produces an image shift $\mathrm{Re}[c'_{11}]$. \textbf{(d)}~Spherical aberration: the outer rays are over-deflected. \textbf{(e)}~Coma: rays are asymmetrically deflected. \textbf{(f)}~Beam tilt combined with spherical aberration produces a coma-like pattern. In panels~\textbf{(c)} and~\textbf{(f)}, the arc above the specimen indicates the beam tilt angle $\mathrm{Re}[t]$ (formally defined in Section~\ref{sec:problem-formulation}). Note that beam tilt, image shift, and coma each have an additional component in the orthogonal direction, not shown here. The ray trajectories follow from a thin-lens model; component distances are simplified for clarity.}
    \label{fig:tilts_shifts_aberrations}
\end{figure}

TEM images are created by directing a beam of electrons onto a specimen and recording the electrons after they have interacted with it. However, accurately guiding the electrons is a non-trivial task. The electron beam is shaped and focused by electromagnetic lenses, which inherently suffer from high sensitivity to environmental conditions and manufacturing inconsistencies. The resulting imperfections in the electron beam are known as aberrations, see \figref\ref{fig:tilts_shifts_aberrations} for a schematic representation of several aberration types. Under the presence of aberrations, the resulting TEM images suffer from reduced contrast, loss of resolution, and blurring.

Fortunately, the aberrations can be compensated by adjusting the electron optical components such as electromagnetic lenses, deflectors, and stigmators. However, this compensation is not straightforward, as the aberrations cannot be measured directly. Additionally, the aberrations drift over time due to thermal effects, mechanical vibrations, and electromagnetic field fluctuations. This limits both image quality and TEM throughput. As a result, fast and accurate estimates of the aberration values are highly desirable.

A well-known principle from optics is that tilting the electron beam in the presence of aberrations causes an apparent image shift in the $x,y$-plane~\citep{Hanszen1971}, revealing information about the underlying aberration coefficients, see \figref\ref{fig:tilts_shifts_aberrations} for an illustration of the aberrations and the tilt-shift interaction. This principle has been employed in several aberration estimation approaches. Classical methods such as the Zemlin tableau analyze diffractograms acquired at different beam tilts~\citep{Zemlin1978}, while more direct approaches estimate aberrations from the tilt-induced image shifts themselves~\citep{Koster1987,Steinecker2000}. However, these methods face significant limitations: diffractogram-based approaches are restricted to specimens that produce suitable diffractograms (typically requiring amorphous regions), limiting their applicability to only a subset of specimen types. Additionally, they are time-consuming. Existing image-shift methods are limited by lengthy image acquisition and processing times that restrict the number of tilts that can be applied. Crucially, none of these approaches model specimen drift explicitly~\citep{Glaeser2011}, making it difficult to separate drift effects from aberration changes.

This paper introduces a method for tilt-based aberration estimation that addresses these limitations. Fast GPU-based image shift tracking~\citep{VanHorssen2020} enables rapid acquisition of many tilt-induced shifts, which are processed in a Kalman filtering (KF) framework that naturally fits the physical model and accounts for aberration drifting. Importantly, the speed and accuracy of the resulting estimates depend on the choice of beam tilt sequence, which motivates the core question addressed in this research: \emph{what sequence of tilts results in the fastest and most accurate aberration estimates?}
The problem of designing measurement sequences to optimize estimation performance has been studied extensively in sensor scheduling and experiment design~\citep{Pukelsheim2006,Atkinson2007}, including adaptations to Kalman filtering for linear discrete-time systems~\citep{Zhang2017}, continuous scheduling parameters~\citep{You2013,Li2015}, and receding-horizon approaches~\citep{Sunberg2016}. However, these works are not tailored to TEM aberration estimation, which involves the freedom to select a continuous, two-dimensional scheduling parameter at every time step and a non-convex but differentiable measurement model.

We address these challenges by formulating the tilt sequence selection as a sensor scheduling problem and proposing a tailor-made solution. Aberrations are estimated using standard Kalman filtering, and the tilt sequence is optimized by minimizing the trace of the predicted covariance. The optimization uses gradients from multiple starting points with receding-horizon optimization. The main contributions are:
\begin{itemize}
  \item A formulation of the tilt-based aberration estimation problem as a sensor scheduling problem, extended to include specimen drift.
  \item Efficient, receding-horizon gradient-based sensor scheduling optimization based on the trace of the predicted covariance matrix in the Kalman filter.
  \item Estimation of specimen-dependent properties using expectation maximization (EM) to tailor the optimization and estimation.
  \item Experimental validation of optimized patterns on a real TEM set-up, including a quantitative comparison with state-of-the-art Zemlin tableau methods using standard image quality metrics.
\end{itemize}

The rest of this paper is organized as follows. Section~\ref{sec:problem-formulation} formulates the aberration estimation problem. Section~\ref{sec:methods} details the methods including the tilt-induced aberration model and tilt pattern optimization. Section~\ref{sec:results} presents results on a real TEM set-up. Section~\ref{sec:conclusion} concludes the paper.

\section{Problem Formulation}
\label{sec:problem-formulation}
\begin{figure}[!t]
    \centering
    \includegraphics[width=60mm]{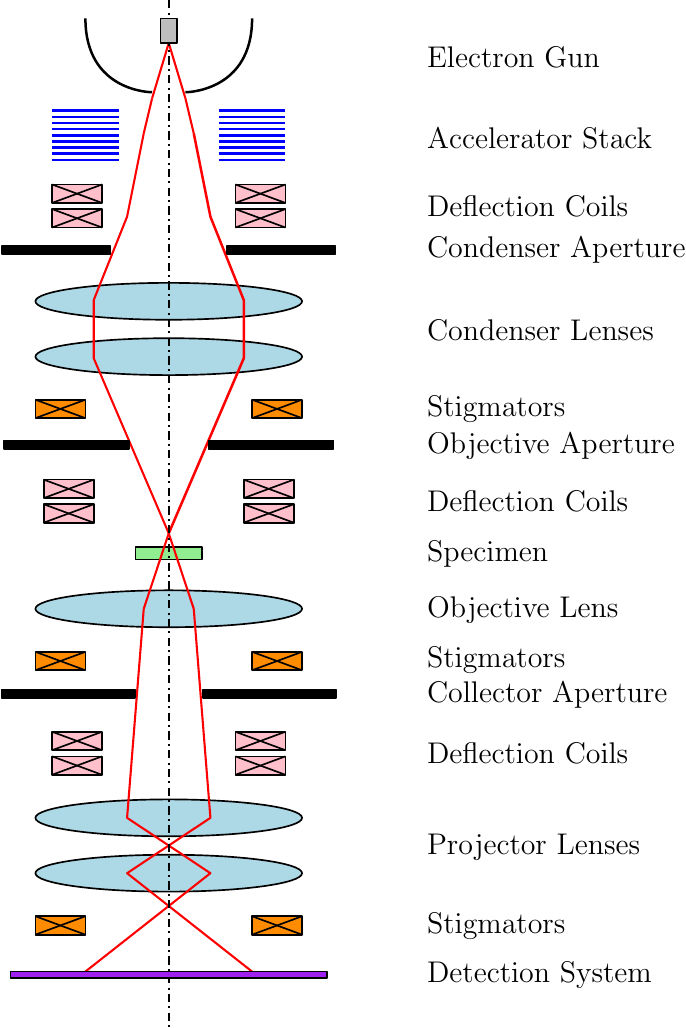}
    \caption{Schematic representation of TEM. An electron gun fires a beam of electrons into a stack of lenses and apertures. Between the lenses, there are stigmators and deflection coils which allow for active compensation of the lens aberrations. The transmitted electrons are collected after interaction with the specimen in a detection system. Note that the width of the electron beam is not drawn to scale. In reality, the beam width is on the order of a few {\AA}ngströms, while the distances between the lenses and specimen are on the order of millimeters. The total column height is typically on the order of a meter.}
    \label{fig:TEM_diagram}
\end{figure}
\begin{figure}[t]
  \centering
  \hspace{-5mm}\includegraphics[width=80mm]{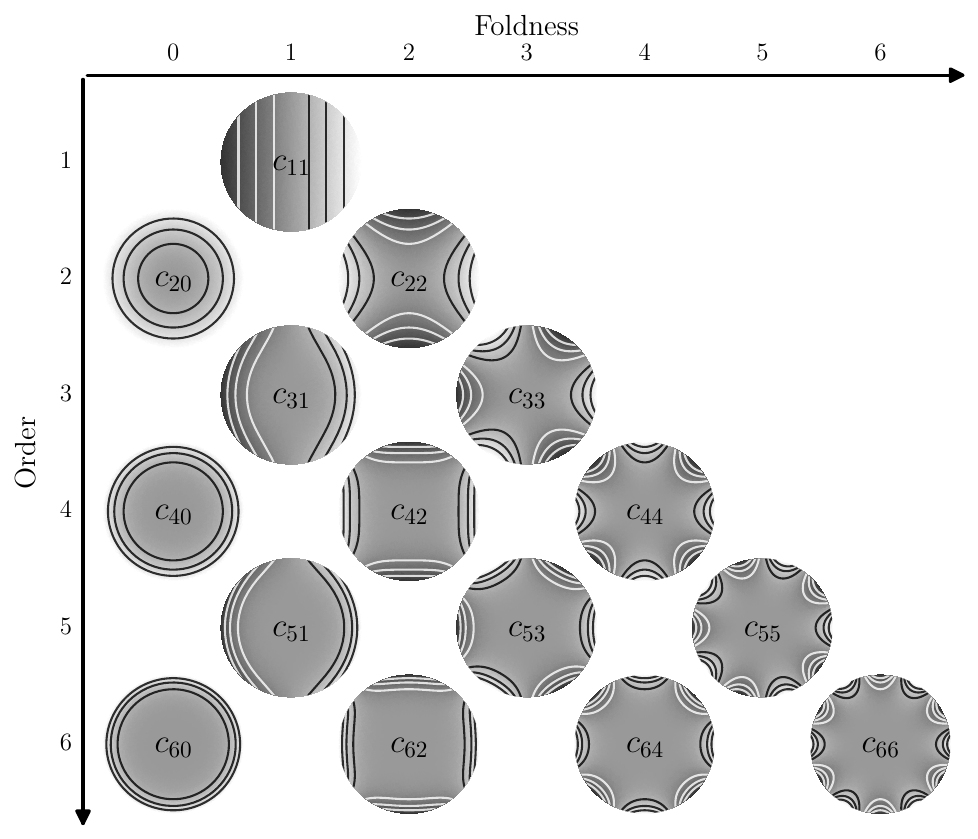}
  \caption{Phase plates of the single wave aberrations in TEM. The columns and rows correspond to the foldness and order of the aberration, respectively. Each phase plate displays the phase shift of the wavefront as a function of the spatial frequency $g \in \mathbb{C}$ in gray scale. Contour lines are plotted in white for positive phase shifts and in black for negative phase shifts. This figure has been adapted from~\citet[Chapter 2]{JuriBarthel2003}.}
  \label{fig:aberration_table}
\end{figure}

TEM forms high-resolution images by transmitting electrons (accelerated to 100-300 keV) through a specimen, see \figref\ref{fig:TEM_diagram}. The transmission imprints information onto the electron wave by altering its phase. This imprint is enlarged by electromagnetic lenses, which project the electron wave onto a detector. The lenses are inherently prone to imperfect focusing due to geometric and electromagnetic field characteristics and finite manufacturing precision. Additionally, environmental conditions such as temperature fluctuations and mechanical vibrations affect lens performance. These imperfections manifest as aberrations, altering the electron wavefront and reducing image resolution. Aberrations are characterized by the contrast transfer function (CTF), which describes how the electron wavefront evolves through the microscope~\citep{JuriBarthel2003,Kirkland2010}. The CTF is defined as
\begin{equation}
  \label{eq:CTF}
  \text{CTF}(g) = E(g)~\text{exp}\left[-i \chi(g)\right],
\end{equation}
where $g \in \mathbb{C}$ is the spatial frequency, $E(g)$ is the envelope function, and $\chi(g)$ is the wave aberration function. The variables are typically defined as complex numbers to represent 2D spatial coordinates.
The envelope function $E(g)$ models coherence effects causing damping across frequencies and is assumed constant. For optimal contrast, $\text{exp}\left[-i \chi(g)\right] \approx 1$ over a broad range of frequencies $g$. The wave aberration function $\chi(g)$ describes the phase shift of the electron wavefront and can be represented as
\begin{equation}
  \label{eq:wave-aberration-function}
  \chi(g) = \sum_{m,n, ~ m-n=2p} \chi_{mn}(g),
\end{equation}
where $m > 0$ and $p, n \geq 0$ are integers such that $m-n=2p$. Each term $\chi_{mn}$ is a degree-$m$ homogeneous polynomial in $\mathrm{Re}[g]$ and $\mathrm{Im}[g]$, and is linear in the aberration coefficient $c_{mn} \in \mathbb{C}$:
\begin{equation}
  \label{eq:wave-aberration-function-bases}
  \chi_{mn}(g, g^*) = \frac{2 \pi}{\lambda}\frac{\lambda^{m}}{m}\,\mathrm{Re}\!\left[c_{mn}\cdot(g^*)^{(m+n)/2} \cdot g^{(m-n)/2}\right],
\end{equation}
where $\lambda$ is the electron beam wavelength, $g^*$ denotes the complex conjugate of $g$, and $c_{mn}$ denotes the aberration coefficient with order $m$ and foldness $n$ (also called multiplicity). The influence of each term can be visualized using phase plates, see \figref\ref{fig:aberration_table}. For a comprehensive overview of aberration types and naming conventions across different notations, we refer to~\citet[Chapter 2.2]{JuriBarthel2003}.

By adjusting the aberration coefficients $c_{mn}$ appropriately, we can control the phase shift and thus the CTF. The coefficients can be adjusted by changing the settings of electron optical components including electromagnetic lenses. However, the coefficients are not directly measurable and must be estimated, which motivates the need for quick and accurate estimation methods.

\subsection{Tilt-induced Aberration Model}
A key physical insight that can be used to estimate the aberrations is that deliberately tilting the electron beam interacts with the aberrations of the microscope. This interaction modifies the effective aberrations, which in turn leads to measurable changes in the image, such as an apparent shifting of the image~\citep{Hanszen1971}.
We can express the effective aberrations as a function of the baseline aberrations and the beam tilt through
{\small
\begin{equation}
  \label{eq:tilt-induced-aberration}
  \begin{aligned}
  c_{(\alpha+\gamma)(\alpha-\gamma)}^{\prime}&=\rho\left(\alpha, \gamma\right) \times(\alpha+\gamma) \sum_{\beta=\alpha}^M \sum_{\delta=\gamma}^{\zeta(M,\beta)} {\Biggl[\binom{\beta}{\alpha}\binom{\delta}{\gamma}\left(t^*\right)^{\beta-\alpha} t^{\delta-\gamma}} \\
  &\quad \times\frac{c_{(\beta+\delta)(\beta-\delta)}}{\beta+\delta}+\binom{\beta}{\gamma}\binom{\delta}{\alpha}\left(t^*\right)^{\delta-\alpha} t^{\beta-\gamma} \frac{c_{(\beta+\delta)(\beta-\delta)}^*}{\beta+\delta}\Biggr],
  \end{aligned}
\end{equation}
}%
where $c_{mn} \in \mathbb{C}$ denotes the baseline aberration with order $m$ and foldness $n$, $c^\prime_{mn} \in \mathbb{C}$ denotes the effective aberration (following the tilt), and $t \in \mathbb{C}$ denotes the beam tilt, where $t^*$ denotes the complex conjugate of $t$. $M$ is the maximum order of baseline aberrations in the model.
The indices $\alpha, \beta, \gamma, \delta$ are summation indices and the function $\zeta(M,\beta)=\mathrm{min}(\beta,M-\beta)$ limits the maximum foldness. The scaling factor $\rho(\alpha, \gamma)$ equals $\frac{1}{2}$ if $\alpha = \gamma$ and $1$ otherwise. Note that the effective aberrations are linear in the baseline aberrations but generally nonlinear in the tilts. A concrete example for $M=2$ is given in~\eqref{eq:tilt-induced-shift}.

Among the effective aberrations in~\eqref{eq:tilt-induced-aberration}, the image shift $c^\prime_{11}$ is of particular interest because it can be directly estimated from TEM images. In practice, this effective shift $c^\prime_{11}$ is measured using a tracking algorithm based on cross-correlations between pairs of images, see e.g.~\citet{VanHorssen2020,Bolme2010}. Tracking works by constructing a so-called template image from one or more of the earlier acquired images. These approaches typically need a number of near-identical images to ensure the template is initialized correctly. In our use case this can be achieved in practice by starting with relatively small initial beam tilts, which ensures small image changes.

The tilt-induced image shift $c^\prime_{11}$ can be linearly related to the baseline aberration coefficients. To see this, we set $(\alpha+\gamma, \alpha-\gamma) = (1,1)$ in~\eqref{eq:tilt-induced-aberration} and evaluate the right-hand side for some maximum order $M$ of baseline aberrations. We then collect the baseline aberration coefficients into a real vector $\vc \in \mathbb{R}^{n_c}$ (where $n_c$ is the number of real-valued aberration coefficients considered). Afterward, we can express the real and imaginary parts of $c^\prime_{11}$ as a linear function of $\vc$:
\begin{equation}
  \vcp = \Psi(\vt) \vc
\end{equation}
with $\vcp = \begin{pmatrix} \real{c^{\prime}_{11}}, \imag{c^{\prime}_{11}} \end{pmatrix}^\top \in \mathbb{R}^2$, $\Psi(\vt) \in \mathbb{R}^{2 \times n_c}$, and $\vt := \begin{pmatrix} \tx, \ty \end{pmatrix}^\top \in \mathbb{R}^2$ the real-valued representation of the beam tilt $t$. For $M=2$, we have $n_c=5$ and obtain
{\small
\begin{equation}
  \label{eq:tilt-induced-shift}
  \underbrace{
  \begin{pmatrix}
    \real{c^{\prime}_{11}} \\
    \imag{c^{\prime}_{11}}
  \end{pmatrix}
  }_{\vcp}
  =
  \underbrace{
  \begin{pmatrix}
    1 & 0 & \tx & \tx & \ty \\ 
    0 & 1 & \ty & -\ty & \tx
  \end{pmatrix}}_{\Psi(\vt)}
  \underbrace{
  \begin{pmatrix}
    \real{c_{11}} \\
    \imag{c_{11}} \\
    c_{20} \\
    \real{c_{22}} \\
    \imag{c_{22}} \\
  \end{pmatrix}
  }_{\vc}.
\end{equation}
}%
From the equation, it is clear how beam tilts $\vt$ reveal the underlying aberration coefficients $\vc$ through measurable shifts $\vcp$. Note that the estimation problem is complicated by the fact that the aberrations drift over time due to changing environmental conditions. Additionally, the particular tilt sequence has a significant impact on estimation speed and accuracy, motivating the search for optimal tilt sequences.

\section{Methods}
\label{sec:methods}
This section presents the methodology for aberration estimation and tilt sequence optimization in TEM. The tilt-induced shift equations with an aberration drifting model admit a linear Kalman filter (KF) formulation, which allows us to estimate the aberrations in a principled way. We then detail the tilt sequence optimization method, which is the main contribution, and develop an approach to estimate specimen-dependent measurement noise covariance using expectation maximization.

\subsection{Dynamic Tilt-Aberration Model}
The tilt-induced shift equations~\eqref{eq:tilt-induced-shift} can be used to predict the image shift in the presence of tilts and aberrations. However, the aberrations drift due to changing environmental conditions. For this reason, we introduce a dynamic model to improve estimation accuracy by relating aberrations to a sequence of measurements. This model includes two key physical effects.

First, the specimen may move with respect to the electron beam, a phenomenon known as specimen drift. This specimen drift is indistinguishable from changes in the image displacement aberration $c_{11}$. We model the drift by treating $c_{11}$ as a time-varying state using a polynomial in time of order $b \in \mathbb{N}$. Since $c_{11} \in \mathbb{C}$ is a complex number representing drift in two dimensions, the polynomial requires $b$ complex states (or equivalently $2b$ real states). Throughout the rest of this paper, we use subscript $k \in \mathbb{N}$ to denote the discrete time index.

Second, the camera axis can in some cases be misaligned with respect to the tilt axis, resulting in relative rotation of tilt and shift coordinates. This is modeled by a state $\phi \in \mathbb{R}$ representing the rotational misalignment. We obtain a linear state-space form
\begin{align}
    x_{k+1} &= A x_k + \xi_k
    \label{eq:state-space-model-dynamics}\\
    y_k &= C(\vt_k)x_k + \varepsilon_k
    \label{eq:state-space-model-observations}
\end{align}
with $x_k \in \mathbb{R}^{d}$ the state and $y_k=\vcp_k \in \mathbb{R}^2$ the measurements, where $d = n_c + 1 + 2b$ and $n_c$ is the number of aberrations considered. The states are normalized to typical values, as aberrations and drift terms differ in magnitude by several orders. This normalization ensures that the filter remains numerically well-conditioned and that the optimization criteria (used in Section~\ref{subsec:optimization}) are not dominated by large-magnitude states. The process noise is $\xi_k \sim \mathcal{N}(0, \Sigma_\xi)$ and the measurement noise is $\varepsilon_k \sim \mathcal{N}(0, \Sigma_\varepsilon)$.
For $b=2$, which results in a second-order polynomial model for specimen drift, we obtain 
{\small
\begin{equation}
  x_k = \left(
    \begin{array}{c}
      \vc_k \\
      \phi_k \\
      \hline
      \real{v_k} \\
      \real{a_k} \\
      \imag{v_k} \\
      \imag{a_k} \\
    \end{array}
  \right),
  \quad
  A =
  \left(
    \begin{array}{>{\centering\arraybackslash}p{1.3cm}|cccc}
      \multirow[c]{5}{*}{$I_{n_c+1}$} 
        & \tau           & \frac{1}{2}\tau^2 & 0   & 0   \\
      & 0             & 0               & \tau & \frac{1}{2}\tau^2 \\
      & 0             & 0               & 0   & 0               \\
      & \multicolumn{4}{c}{\vdots}                       \\
      & 0             & 0               & 0   & 0               \\ 
      \hline
      \multirow[c]{4}{*}{$0$} 
        & 1             & \tau             & 0   & 0   \\
      & 0             & 1               & 0   & 0   \\
      & 0             & 0               & 1   & \tau \\
      & 0             & 0               & 0   & 1   
    \end{array}
  \right),
\end{equation}
}%
and
\begin{equation}
  C(\vt_k) = \left(
    \begin{array}{cc|cccc}
      \multirow[c]{2}{*}{$\Psi\left(\vt_k\right)$} & -\tyk & 0 & 0 & 0 & 0 \\
      & \txk & 0 & 0 & 0 & 0 
    \end{array}
  \right),
\end{equation}
where $v_k \in \mathbb{C}$ and $a_k \in \mathbb{C}$ are the drift velocity and acceleration at time index $k$, with $\real{v_k}$ and $\imag{v_k}$ representing the velocity components in the $x$ and $y$ directions, respectively, and similarly for $\real{a_k}$ and $\imag{a_k}$. $\tau$ denotes the sampling time. Here, $\vc_k$ is the time-varying counterpart of the aberration vector $\vc$ introduced in~\eqref{eq:tilt-induced-shift}, where the subscript $k$ reflects that aberrations are now treated as a dynamic state. Essentially, the aberrations $\vc_k$ (elements of $x_k$) remain constant (in the absence of process noise $\xi_k$), except for the shift aberrations $\real{c_{11}}$ and $\imag{c_{11}}$, which are modeled as second-order polynomials in time. Note that the model can be easily extended to include polynomial drifting up to arbitrary orders of other aberrations, which would come at the cost of a more complex estimation problem. We assume the normalized initial state $x_0 \sim \mathcal{N}(0, I_{d})$.

\subsection{Kalman filtering for Tilt-based Aberration Estimation}
To estimate the state $x_k$ using the model~\eqref{eq:state-space-model-dynamics},~\eqref{eq:state-space-model-observations} based on a sequence of measurements $y_0, \ldots, y_k$, we employ the Kalman filter. The variable $\tilde{x}_{k\mid l}$ denotes the estimate of state $x_k$ given measurements up to time $l$, i.e., $\tilde{x}_{k\mid l} := E[x_k \mid y_0,\vt_0 \ldots, y_l,\vt_l]$. Similarly, $P_{k\mid l}$ is the estimation error covariance, $E[(x_k - \tilde{x}_{k\mid l})(x_k - \tilde{x}_{k\mid l})^T \mid y_0,\vt_0 \ldots, y_l,\vt_l]$. The filter is initialized with $\tilde{x}_{0\mid-1} = 0$ and $P_{0\mid-1} = I_{d}$. The standard KF recursively computes estimates through update and prediction steps (see~\citet{Kalman1960} for details).

The estimation could alternatively be approached using batch Linear Least Squares (LLS). For linear Gaussian systems, the KF and LLS yield equivalent estimates when the same prior knowledge is incorporated (i.e., in LLS through regularization and weighting). However, we opt for the KF framework because it naturally models aberration drift through state derivatives and process noise $\xi_k$, and intuitively incorporates prior beliefs via $\tilde{x}_{0\mid-1}$, $P_{0\mid-1}$, $\Sigma_\varepsilon$, and $\Sigma_\xi$. While these aspects can also be addressed in LLS, the Bayesian recursive nature of the KF handles them more naturally.

A key result is that the estimation error covariance $P$ evolves independently of actual measurements~\citep{Kalman1960,Atkinson2007}, enabling offline optimization. If $A$ is invertible, we can batch all measurements and use $N$-step predicted covariances:
\begin{equation}
    P_{N-1\mid -1} = A^{(N-1)} P_{0\mid -1} (A^\top)^{(N-1)} + \sum_{i=0}^{N-2} A^i \Sigma_\xi (A^\top)^i,
    \label{eq:N-step-predicted-covariance}
\end{equation}
\begin{equation}
\begin{aligned}
    P_{N-1\mid N-1} &= P_{N-1\mid -1} - P_{N-1\mid -1}\bar{C}^\top(\mathcal{T}_N)S^{-1}(\mathcal{T}_N) \\
    & \qquad \qquad\qquad\qquad \times \bar{C}(\mathcal{T}_N)P_{N-1\mid -1},
  \end{aligned}
  \label{eq:N-step-corrected-covariance}
\end{equation}
with 
\begin{equation}
  \label{eq:lifted-observation-matrix}
  \bar{C}(\mathcal{T}_N) = \begin{bmatrix}
    C(\vt_0)A^{-(N-1)} \\
    C(\vt_1)A^{-(N-2)} \\
    \vdots \\
    C(\vt_{N-1})
  \end{bmatrix},
\end{equation}
and
\begin{equation}
  \label{eq:innovation-covariance}
  S(\mathcal{T}_N) = \bar{C}(\mathcal{T}_N) P_{N-1\mid -1} \bar{C}^\top(\mathcal{T}_N) + I_N \otimes \Sigma_\varepsilon,
\end{equation}
where $\otimes$ denotes the Kronecker product, and $\mathcal{T}_N:= \left\{\vt_i\right\}_{i=0}^{N-1}$ is the full tilt sequence.

We note that the batched form~\eqref{eq:N-step-corrected-covariance} is a standard equivalent reformulation of the recursive Kalman filter~\citep{Simon2006}, expressing the $N$-step posterior covariance directly in terms of the prior $P_{0\mid -1}$ and the full tilt sequence $\mathcal{T}_N$. This batched KF formulation shares structural similarities with moving-horizon estimation (MHE)~\citep{Rao2003} and receding-horizon FIR filtering~\citep{Kwon2002}, which also process a batch of measurements. In our setting, however, we process the full measurement history from the start, whereas MHE and FIR typically use a sliding window of recent measurements. Additionally, in this work we use the batch form primarily for offline covariance prediction rather than real-time state estimation.

Given~\eqref{eq:N-step-predicted-covariance}--\eqref{eq:innovation-covariance}, we can compare tilt sequences in terms of their impact on estimation error covariance, which is a measure of uncertainty on the aberration values. The ability to predict the estimation error covariance enables the design of sequences that minimize this uncertainty.

\subsection{Optimization of Tilt Sequences}
\label{subsec:optimization}
Sensor scheduling involves actively deciding sensor configurations over time to optimize an objective, typically related to estimation performance. By choosing the beam tilt $\vt_k$, we change the matrix $C(\vt_k)$ in~\eqref{eq:state-space-model-observations}. The goal is to determine the $N$-step tilt sequence $\mathcal{T}_N$ that minimizes aberration estimation uncertainty using the predicted posterior covariance. We assess tilt pattern performance by taking a weighted trace of the predicted error covariance:
\begin{equation}
  \label{eq:optimization_problem}
  \begin{aligned}
    \min_{\mathcal{T}_N} \quad & \text{tr}\left(W P_{N-1\mid N-1}\right), \\
  \text{s.t.} \quad & ~\eqref{eq:N-step-predicted-covariance},~\eqref{eq:N-step-corrected-covariance},~\eqref{eq:lifted-observation-matrix},~\eqref{eq:innovation-covariance}, \\
  & \lVert \vt_k \rVert_2 \leq t^{\max}_k, \quad k \in \{0,1,\ldots,N-1\},
  \end{aligned}
\end{equation}
where the tilt magnitude constraint will be discussed later. By taking the covariance trace as the cost, we essentially minimize the sum of variances of each of the aberration estimates (for $W=I_{d}$), known as A-optimality~\citep{Atkinson2007}. Other common optimality criteria in experiment design include D-optimality (minimizing the determinant) and E-optimality (minimizing the largest eigenvalue), which can be interpreted as minimizing the volume of the uncertainty ellipsoid and minimizing the worst-case variance, respectively. We choose A-optimality as it provides an appropriate overall uncertainty measure and is computationally convenient, though the results extend to other criteria. In batch experiment design using LLS, criteria are frequently derived from the Fisher Information Matrix (FIM), which is the inverse of the covariance matrix. In the KF framework, working directly with the posterior covariance is a natural approach.

The matrix $W \in \mathbb{R}^{d \times d}$ is a positive semi-definite weighting matrix ($W \succeq 0$) that can be used to prioritize estimation accuracy of certain aberrations over others. This weighting is applied \emph{after} state normalization, ensuring users express preferences among comparable state values.

\begin{figure}[t]
  \centering
  \includegraphics[width=65mm]{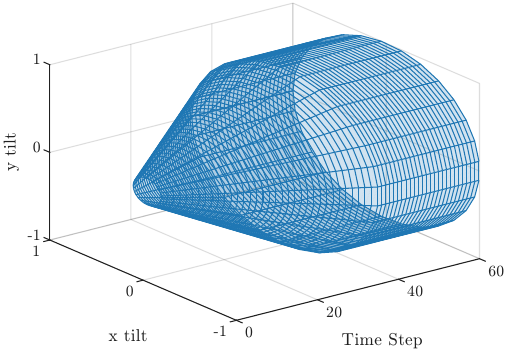}
  \caption{Example of a user-defined tilt constraint $t^{\max}_k$ as a function of time $k$. The red lines indicate the boundaries of the feasible set. Note that the tilt values are normalized.}
  \label{fig:tilt_constraints}
\end{figure}
We constrain the tilt magnitude to be less than a user-defined scalar bound $t^{\max}_k \in \mathbb{R}_{>0}$, which varies with time index $k$. This allows us to address image tracker template initialization by limiting tilts at the beginning and increasing them as tracker accuracy improves. Additionally, the tilt magnitude must be upper-bounded since~\eqref{eq:tilt-induced-aberration} is only valid for small tilts. \figref\ref{fig:tilt_constraints} shows an example of the tilt magnitude constraint as a function of time.

Several aspects make this sensor scheduling problem distinct from standard sensor scheduling problems:
\begin{enumerate}
\item The sensor scheduling parameter is a \emph{continuous}, two-dimensional vector that can be selected at \emph{every timestep};
\item The cost function $\text{tr}( WP_{N-1 \mid N-1} )$ is \emph{differentiable} with respect to $\mathcal{T}_N$, enabling efficient gradient-based optimization;
\item Despite differentiability, the cost function is \emph{non-convex} in the design variables, implying multiple local minima may exist;
\item The feasible set is convex and can be represented as a simple box constraint in polar coordinates.
\end{enumerate}
We now detail how to solve the optimization problem~\eqref{eq:optimization_problem} using these properties.

The first key insight is that we can take an analytical gradient with respect to the design variables, enabling efficient gradient-based algorithms. Using the identities
\begin{equation}
  \frac{\partial (\text{tr}\left(X(y)\right))}{\partial y} = \text{tr}\left(\frac{\partial X(y)}{\partial y}\right),
\end{equation}
and
\begin{equation}
  \frac{\partial (X^{-1}(y))}{\partial y} = -X^{-1}(y) \frac{\partial X(y)}{\partial y} X^{-1}(y),
\end{equation}
which hold for any vector $y$ and invertible matrix-valued function $X(y)$, we obtain
{\small
\begin{equation}
  \begin{aligned}
    \frac{\partial P_{N-1\mid N-1}}{\partial \mathcal{T}_N} &= -P_{N-1\mid -1}
    \begin{bmatrix}
      S^{-1}(\mathcal{T}_N)\bar{C}(\mathcal{T}_N)\\
      \frac{\partial \bar{C}(\mathcal{T}_N)}{\partial \mathcal{T}_N}
    \end{bmatrix}^\top
    \begin{bmatrix}
      \Omega(\mathcal{T}_N) & 1\\
      1 & 0
    \end{bmatrix} \\
    & \qquad \qquad \quad \times 
    \begin{bmatrix}
      S^{-1}(\mathcal{T}_N) \bar{C}(\mathcal{T}_N) \\
      \frac{\partial \bar{C}(\mathcal{T}_N)}{\partial \mathcal{T}_N}
    \end{bmatrix}
    P_{N-1\mid -1},
  \end{aligned}
\end{equation}
}%
in which
\begin{equation}
  \begin{aligned}
  \Omega(\mathcal{T}_N) &= -\biggl(\frac{\partial \bar{C}(\mathcal{T}_N)}{\partial \mathcal{T}_N} P_{N-1\mid -1} \bar{C}^\top(\mathcal{T}_N) \\
  & \qquad\quad + \bar{C}(\mathcal{T}_N) P_{N-1\mid -1} \frac{\partial \bar{C}^\top(\mathcal{T}_N)}{\partial \mathcal{T}_N}\biggr),
  \end{aligned}
\end{equation}
and
\begin{equation}
  \frac{\partial\bar{C}(\mathcal{T}_N)}{\partial \mathcal{T}_N} =
  \begin{bmatrix}
    \frac{\partial C(\vt_0)}{\partial \mathcal{T}_N}A^{-(N-1)} \\
    \frac{\partial C(\vt_1)}{\partial \mathcal{T}_N}A^{-(N-2)} \\
    \vdots \\
    \frac{\partial C(\vt_{N-1})}{\partial \mathcal{T}_N}
  \end{bmatrix}.
\end{equation}
The availability of an analytical gradient is preferred over numerical approximations for efficiency and accuracy. This analytical gradient can be used within a standard constrained optimization algorithm (e.g., a projected gradient method or an interior-point method) that explicitly handles the constraints. To do so, we transform each tilt $\vt_k$ into polar coordinates $(r_k, \psi_k)$, where $r_k$ is the magnitude and $\psi_k$ is the angle. The magnitude constraints $\lVert \vt_k \rVert_2 \leq t^{\max}_k$ then become simple box constraints $0 \leq r_k \leq t^{\max}_k$, $0 \leq \psi_k < 2 \pi$, which are readily incorporated into gradient-based solvers.

\begin{figure}[t]
  \centering
  \includegraphics[width=80mm]{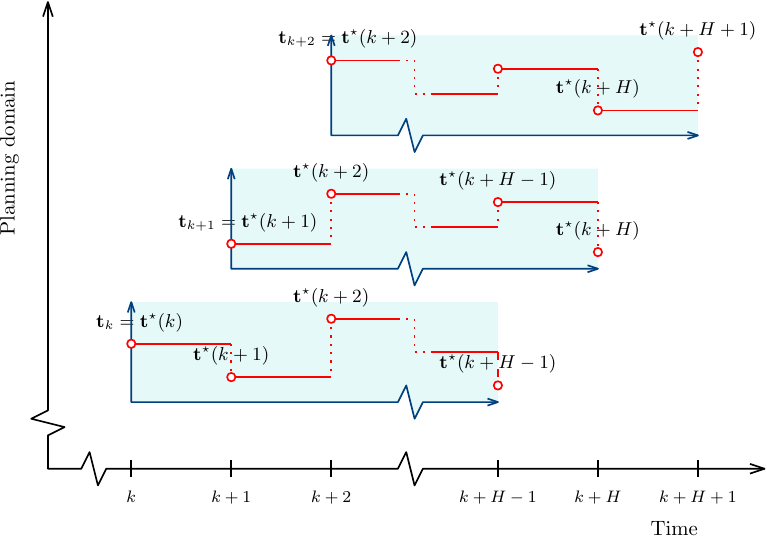}
  \caption{Schematic representation of receding-horizon optimization. Starting at time $k$, we solve an optimization problem with a reduced horizon $H$ to find the optimal tilts $\left\{\vt^\star(i)\right\}_{i=k}^{k+H-1}$. We then store the first tilt $\vt_k = \vt^\star(k)$, move up one time step in the planning domain and solve the $k+1$-th optimization problem with the updated predicted covariance $P_{k+1\mid k}$. We can use the previous solution $\left\{\vt^\star(i)\right\}_{i=k}^{k+H-1}$ as a warm start for the next optimization problem. This process is repeated until we reach the end of the tilt sequence.}
  \label{fig:receding_horizon}
\end{figure}
Next, since the number of design variables scales with the tilt pattern length (i.e., $2N$ scalar variables), we use receding-horizon optimization (RHO) with a horizon $H<N$ to solve a sequence of reduced-size problems. At each current time $k$, we solve:
\begin{equation}
  \label{eq:receding_horizon_optimization}
  \begin{aligned}
    \min_{\left\{\vt(i)\right\}_{i=k}^{k+H-1}} \quad & \text{tr}\left(W P_{k+H-1\mid k+H-1}\right), \\
  \text{s.t.} \quad
  & \lVert \vt(i) \rVert_2 \leq t^{\max}_i, \quad i \in \{k,k+1,\ldots,k+H-1\}, \\
  & P_{k\mid k-1} \text{ given},
  \end{aligned}
\end{equation}
here, $\left\{\vt(i)\right\}_{i=k}^{k+H-1}$ denotes the design variables in the reduced-horizon tilt sequence. After solving and obtaining the optimized solution $\left\{\vt^\star(i)\right\}_{i=k}^{k+H-1}$, we store the first element $\vt_k=\vt^\star(k)$, calculate $P_{k+1 \mid k}$, and repeat with $k \leftarrow k+1$ until $k=N$ (set $H = \min(H,N-k)$). This is illustrated in \figref\ref{fig:receding_horizon}. RHO, even with $H=1$, can provide near-optimal performance~\citep{Mayne2000a} while reducing design variables from $2N$ to $2H$.

Since the cost function is non-convex, we employ a \emph{multi-start strategy}, solving the RHO problem from many randomized initial feasible sequences and selecting the best solution. Starting from $k=1$, we introduce \emph{warm starts} by using the best previous RHO solutions as a subset of initial sequences. We use 1000 starting points total for all optimizations, with 100 warm starts when $k \geq 1$.

Note that the solution is sensitive to system parameters $P_{0\mid -1}$, $\Sigma_\xi$, and $\Sigma_\varepsilon$. While the first two can be chosen based on expert knowledge and historical data, the measurement noise covariance $\Sigma_\varepsilon$ can vary significantly across specimens and imaging conditions. For this reason, we propose a method to estimate it which is detailed in the next section.

\subsection{Estimating specimen-dependent measurement noise\\covariance}
\label{subsec:EM}
Different specimen types and imaging conditions such as field of view and defocus affect the amount of contrast and distinguishable features within the image. Since image tracker accuracy depends on these visual characteristics, different specimens and conditions typically result in different measurement noise covariances $\Sigma_\varepsilon$.

To estimate the specimen-dependent $\Sigma_\varepsilon$, we employ expectation maximization, alternating between Kalman smoothing (E-step) and updating the noise covariance estimate (M-step). In the M-step, we update parameters by maximizing the expected complete-data log-likelihood~\citep{Gibson2005}. For the measurement noise covariance $\Sigma_\varepsilon$, the update is given by
\begin{equation}
  \begin{aligned}
  \label{eq:m_step}
  \Sigma_\varepsilon &= \frac{1}{N} \sum_{k=0}^{N-1} \left( y_k - C(\vt_k)\tilde{x}_{k \mid N-1} \right) \left( y_k - C(\vt_k)\tilde{x}_{k \mid N-1} \right)^\top \\
  & \qquad\qquad\qquad + C(\vt_k) P_{k\mid N-1} C(\vt_k)^\top,
\end{aligned}
\end{equation}
where $\tilde{x}_{k \mid N-1}$ and $P_{k\mid N-1}$ are the smoothed state estimate and covariance, respectively, obtained from the Kalman smoother~\citep{Gibson2005}. The EM algorithm is guaranteed to converge to a local maximum of the likelihood. It is advisable to run it multiple times with different initializations based on expert knowledge or rough estimates, and select the solution with the highest likelihood. This leads to the following adaptive filtering procedure.
\begin{procedure}
\upshape
  \emph{Tilt-based aberration estimation and correction with specimen-dependent tilt pattern optimization}
  \label{proc:aberration_estimation}
\begin{enumerate}
  \item Perform several tilt experiments with a general tilt pattern on the TEM and collect shift estimates.
  \item Run the EM algorithm, alternating E-step and M-step~\eqref{eq:m_step}, until convergence to estimate $\Sigma_\varepsilon$.
  \item Determine weighting matrix $W$, tilt pattern length $N$, and horizon length $H$, then generate an optimized tilt pattern using the method in Section~\ref{subsec:optimization}.
  \item Apply the optimized tilt pattern to the TEM and gather shift measurements.
  \item Use the Kalman filter to estimate the aberrations.
  \item Compensate the aberrations by adjusting the electromagnetic fields.
  \item Repeat Steps 4-6 as necessary.
\end{enumerate}
\end{procedure}

\section{Results}
\label{sec:results}
We present tilt pattern optimization results and experimental validation on a Thermo Fisher Scientific Krios TEM in Eindhoven, the Netherlands, following Procedure~\ref{proc:aberration_estimation}. A polycrystalline gold foil specimen is used for optimization and validation, as it can handle high electron doses, allowing repeated application of different tilt patterns without damage.

We consider aberrations up to order $M=4$ (14 coefficients). The specimen drift is modeled using a second-order polynomial ($b=2$) for both $x$ and $y$ directions, plus image rotation, bringing the total number of states to $d=19$.

\subsection{Optimized Tilt Patterns}
\begin{figure}[t]
  \centering
  \includegraphics[width=85mm]{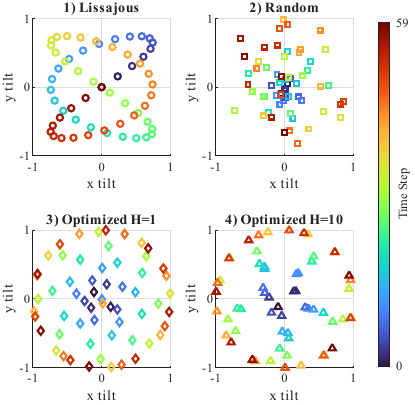}
  \caption{Comparison of different normalized tilt patterns. The patterns are: 1) Lissajous curve with frequency ratio 3:2, 2) completely randomized feasible pattern, 3) optimized pattern with $H=1$, and 4) optimized pattern with $H=10$. The color of each marker indicates the time step $k$, progressing from dark blue ($k=0$) to dark red ($k=59$).}
  \label{fig:tilt_patterns}
\end{figure}
\begin{figure}[t]
  \centering
  \includegraphics[width=85mm]{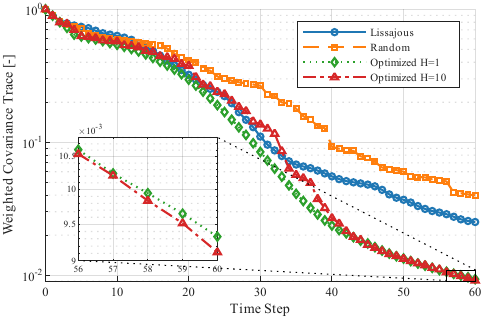}
  \caption{Weighted posterior covariance trace ($\text{tr}(WP_{k\mid k})$) at each time index for selection of tilt patterns. In this figure, the estimation error of every normalized state $x_k$ is weighted equally using weight matrix $W=\frac{1}{d}I_{d}$. The zoomed-in plot on the bottom left focuses on the two optimized patterns at the final time steps, where the greedy pattern 3) very slightly underperforms the 10-step optimized pattern 4).}
  \label{fig:covariance_trace_comparison}
\end{figure}
We apply the optimization method from Section~\ref{sec:methods} to find optimal tilt patterns for TEM aberration estimation. The measurement noise covariance $\Sigma_\varepsilon$ for the gold foil is estimated using the EM algorithm described in the previous section. We use the interior-point algorithm to find local minimizers of~\eqref{eq:receding_horizon_optimization} with tilt sequence length $N=60$. Four tilt patterns are compared in \figref\ref{fig:tilt_patterns}:
\begin{enumerate}
  \item A Lissajous curve with a frequency ratio of 3:2;
  \item A randomized pattern with tilts sampled uniformly from the time-dependent feasible set (see \figref\ref{fig:tilt_constraints});
  \item An optimized pattern generated using gradient-based multi-start optimization with RHO, a horizon of $H=1$ (greedy or myopic), and a weighting matrix $W=\frac{1}{d}I_{d}$;
  \item An optimized pattern generated using gradient-based multi-start optimization with RHO, a horizon of $H=10$, and a weighting matrix $W=\frac{1}{d}I_{d}$.
\end{enumerate}

Patterns 1--2 are not optimized and are therefore expected to have poor performance in terms of the cost criterion. Pattern 3 (greedy optimization, $H=1$) typically trades some performance for computational speed. Pattern 4 is optimized with $H=10$ and is expected to perform well across most aberration types. The computation times for different optimization configurations are summarized in \tabref\ref{tab:optimization_computation_times}.
\begin{table}[t]
  \centering
  \caption{Approximate wall-clock times for the optimization of a 60-tilt sequence with different horizon lengths. Times measured in MATLAB on a laptop computer; actual times may vary with hardware and system load. Note that the optimization is performed offline and does not affect the real-time estimation process. $^{[1]}$These settings correspond to pattern 3) in \figref\ref{fig:tilt_patterns}. $^{[2]}$These settings correspond to pattern 4) in \figref\ref{fig:tilt_patterns}.}
  \label{tab:optimization_computation_times}
  \begin{tabular}{llll}
    \toprule
    \textbf{Horizon} & \textbf{Number of}& \textbf{Comp. Time} & \textbf{Cov. Trace} \\
    \textbf{length $H$}& \textbf{Multi-starts} & \textbf{(seconds)} & $\text{tr}(WP_{N-1 \mid N-1})$\\
    \midrule
    1 & 10 & 37.1 & 1.06e-2\\
    1 & 100 & 117.6 & 9.60e-3\\
    1$^{[1]}$ & 1000 & 485.6 & 9.33e-3\\
    2 & 100 & 140.3 & 9.32e-3\\
    5 & 1000 & 2650.8 & 9.63e-3\\
    10 & 100 & 912.4 & 9.32e-3\\
    10$^{[2]}$ & 1000 & 7643.2 & 9.12e-3\\
    \bottomrule
  \end{tabular}
\end{table}

\figref\ref{fig:covariance_trace_comparison} shows the trace of the weighted estimation error covariance $WP_{k\mid k}$ as a function of time index $k$. The 10-step optimized pattern 4) is expected to have the best performance in terms of the equally-weighted cost criterion. Interestingly, the greedy pattern 3) outperforms all other patterns at the beginning and middle of the estimation process, but falls behind pattern 4) slightly towards the end. This behavior is consistent with the fact that RHO is suboptimal by design: only the first tilt of each optimized horizon is applied, which favors immediately informative tilts. Note that for the greedy pattern, all intermediate patterns of lengths $1$ to $N$ are available without re-optimization, which is a practical advantage. While all patterns significantly reduce the estimates' covariance over time, the optimized patterns typically achieve the same performance in a much shorter time span. For example, the final covariance trace of the random pattern 2) is already reached after approximately 35 time steps with the optimized pattern 3), which represents a reduction of 25 time steps or 42\%.

\subsection{Experimental Validation}
\begin{figure*}[t]
  \centering
  \includegraphics[width=180mm]{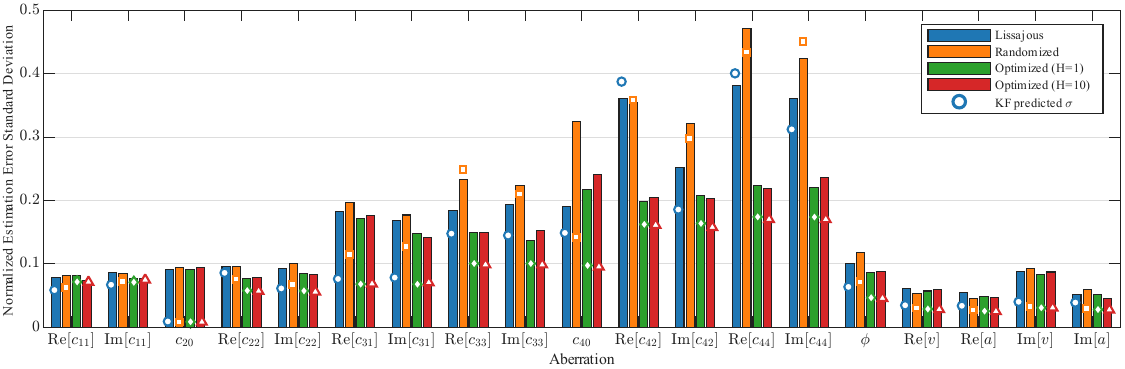}
  \caption{Comparison of the expected and realized standard deviation in aberration estimates for different tilt patterns applied to a gold specimen. The markers indicate the state estimation error predicted by the Kalman filter (i.e., the square root of the diagonal elements of $P_{N-1 \mid N-1}$), while the bars indicate the realized standard deviation in aberration estimates across many experiments. The different colors indicate different tilt patterns, which are the same as those shown in \figref\ref{fig:tilt_patterns}.}
  \label{fig:variance_comparison}
\end{figure*}
To assess the practical performance of the estimation method and the tilt patterns, we perform over 2000 trials on an actual TEM set-up at different specimen locations. At each location, we test the tilt patterns from \figref\ref{fig:tilt_patterns} in randomized order to avoid bias from time-dependent effects.

To validate aberration estimation performance, we compare each experiment's estimated aberrations to the mean aberration estimates from experiments with similar timings and locations. This local mean can be viewed as a ground truth for the aberration values at that time and location. While this metric is sensitive to imperfect modeling, it demonstrates the variance in aberration estimates and the consistency of different tilt patterns.

We compare the experimentally realized variance in aberration estimates to the expected state estimate variance, given by the diagonal elements of $P_{N-1 \mid N-1}$ in the KF, in \figref\ref{fig:variance_comparison}. Several conclusions can be drawn. First, the expected state estimate variance from the KF quite closely resembles the realized variance in aberration estimates, signaling that the tilt-induced shift model is relatively accurate. Typically, realized variances are slightly higher than predicted, likely due to undermodeling of aberrations beyond $M=4$ and non-polynomial drift. The defocus aberration ($c_{20}$) is notably different in this aspect, likely because defocus is retuned at different points during the experiments, violating the constant-aberration assumption. Second, optimized patterns generally outperform non-optimized patterns in terms of realized estimation variance across all aberration types. Third, the greedy pattern ($H=1$) achieves comparable performance to the $H=10$ pattern. In practice, the greedy pattern is preferred due to its computational efficiency and the availability of intermediate-length patterns without re-optimization.

Having validated the estimation performance, we next investigate whether the improved aberration estimates translate into measurable improvements in image quality.

\subsection{Image Quality Comparison}
\label{subsec:image-quality}
To quantitatively evaluate the practical benefit of the proposed method, we compare it against a standard Zemlin tableau~\citep{Zemlin1978} implementation in a Monte Carlo experiment on two specimen types: an amorphous carbon film and a polycrystalline gold specimen. In each trial, the microscope is intentionally set to a realistic aberrated state by perturbing defocus~$c_{20}$, twofold astigmatism~$c_{22}$, and axial coma~$c_{31}$, and a \emph{before} image is captured. Then, one of the two aberration correction methods is applied for several iterations of estimation and correction of these three aberrations, after which an \emph{after} image is captured. For the proposed method, we use the optimized tilt pattern with $H=10$ (pattern~4 in \figref\ref{fig:tilt_patterns}).
To ensure a fair comparison, the correction applied by the first method is reset before applying the second, so that both methods start from approximately the same aberrated state. This procedure is repeated 100 times for both methods and both specimen types.

We assess image quality using the spectral signal-to-noise ratio (SSNR)~\citep{Unser1987}, which quantifies the ratio of signal power to noise power as a function of spatial frequency. For each condition (before correction, after Zemlin correction, and after the proposed correction), the 100 images across trials form a single stack. The SSNR is then computed from this stack. The reproducible content across images constitutes the signal, while the image-to-image variability constitutes the noise. This noise includes contributions from residual aberrations (which vary between trials and which we aim to minimize) as well as from detector noise (which remains constant across methods). Radially averaging the resulting 2D spectra yields 1D SSNR curves as functions of the scalar spatial frequency~$g$.

The resulting SSNR curves show at which spatial frequencies the imaging performance improves after correction. To summarize image quality in a single number, we integrate the signal power spectrum (the numerator of the SSNR~\citep{Unser1987}) and the noise power spectrum (its denominator) separately over all spatial frequencies, and take their ratio. This yields a single unitless signal-to-noise ratio (SNR) that captures the overall image quality. The SNR values are reported in \tabref\ref{tab:cumulative_sps_values}.

\begin{table}[t]
  \centering
  \caption{SNR values (cumulative signal power divided by cumulative noise power) for both methods and both specimen types (higher is better). This unitless metric summarizes overall image quality. The proposed method achieves higher SNR on both specimen types. The difference in values across specimen types is caused by the fact that amorphous carbon and polycrystalline gold have different contrast.}
  \label{tab:cumulative_sps_values}
  \begin{tabular}{lccc}
    \toprule
    \textbf{Specimen Type} & \textbf{Before} & \textbf{Zemlin} & \textbf{Proposed} \\
    \midrule
    Amorphous carbon & 0.168 & 0.169 & 0.198\\
    Polycrystalline gold & 0.716 & 0.719 & 4.372 \\
    \bottomrule
  \end{tabular}
\end{table}

\begin{figure}[t]
  \centering
  \includegraphics[width=85mm]{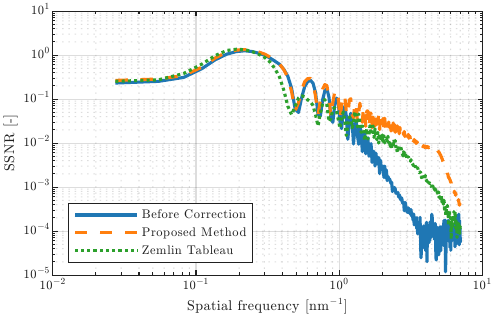}
  \caption{SSNR as a function of spatial frequency for the amorphous carbon specimen, computed from the 100 Monte Carlo trial images. Both the Zemlin tableau method and the proposed method improve the SSNR relative to the uncorrected baseline at every frequency, with the proposed method achieving consistently higher values.}
  \label{fig:cumulative_sps_amorphous}
\end{figure}

\begin{figure}[t]
  \centering
  \includegraphics[width=85mm]{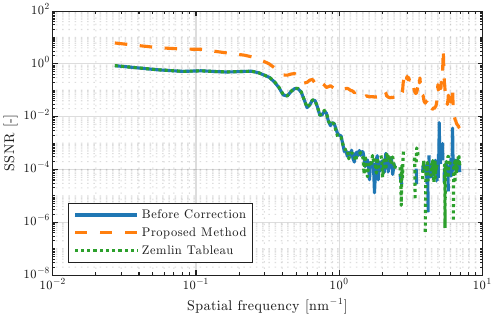}
  \caption{SSNR as a function of spatial frequency for the polycrystalline gold specimen, computed from the 100 Monte Carlo trial images. The proposed method substantially improves the SSNR at every frequency on this specimen type. The Zemlin tableau fails on polycrystalline specimens and therefore does not improve the SSNR.}
  \label{fig:cumulative_sps_gold}
\end{figure}

\figref\ref{fig:cumulative_sps_amorphous} shows the SSNR comparison for the amorphous carbon specimen. Both methods substantially increase the SSNR relative to the uncorrected baseline across all spatial frequencies. The proposed method consistently achieves higher values than the Zemlin tableau, demonstrating that optimized tilt-based estimation yields accurate aberration correction even on amorphous specimens, the specimen type for which the Zemlin tableau is specifically designed. The observed improvement over the Zemlin tableau on this specimen type is modest and may partly reflect the particular Zemlin implementation tested. The key observation is that the proposed method achieves at least comparable aberration correction quality on amorphous specimens while also extending to non-amorphous specimens, as shown next.

\begin{figure}[t]
  \centering
  \includegraphics[width=35mm]{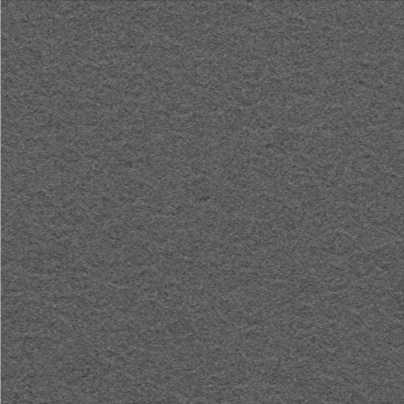}
  \hspace{0.5cm}
  \includegraphics[width=35mm]{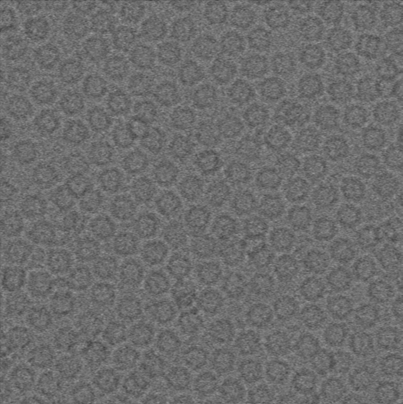}
  \caption{Apoferritin specimen imaged before (left) and after (right) aberration correction using the proposed tilt-based method. In the uncorrected image, structural features are difficult to discern due to suboptimal contrast transfer. After correction, the characteristic ring-like molecular structure of apoferritin becomes clearly visible, indicating improved imaging conditions.}
  \label{fig:image_quality_comparison}
\end{figure}

\figref\ref{fig:cumulative_sps_gold} shows the corresponding results for the polycrystalline gold specimen. Here, the advantage of the proposed method is substantially larger. This is expected: diffractogram-based methods such as the Zemlin tableau rely on Thon rings produced by amorphous regions, which are absent in polycrystalline specimens, and as a result the Zemlin tableau correction completely fails.
The proposed method, relying solely on image shift measurements, is not subject to this limitation. This specimen-type independence is particularly relevant in life-science workflows, where aberrations are typically corrected on a polycrystalline gold substrate before moving the beam to the vitrified biological specimen of interest, minimizing electron dose on dose-sensitive material.
\figref\ref{fig:image_quality_comparison} illustrates this for an apoferritin specimen, a protein commonly studied in cryo-electron microscopy. After correcting aberrations on a gold region using the proposed method, the characteristic ring-like molecular structure of apoferritin becomes clearly visible, confirming that the aberrations have been sufficiently reduced to resolve the specimen's features.

\section{Conclusion}
\label{sec:conclusion}
Image-shift based aberration estimation in TEM has historically been limited by sensitivity to specimen drift and slow image acquisition and processing. This paper addresses both of these barriers. Fast GPU-based image shift tracking accelerates the measurement process, enabling the application of more tilts within a given time frame. The main contribution of this paper is a method to optimize the sequence of applied tilts to maximize information gain. To this end, we formulate tilt-based aberration estimation as a sensor scheduling problem within a Kalman filtering framework that leverages the physical relationship between beam tilts and image shifts to estimate multiple aberration coefficients simultaneously. A key element of the framework is that it explicitly accounts for the temporal evolution of aberration coefficients, including specimen drift. The tilt sequence is optimized offline using an A-optimality criterion through gradient-based receding-horizon optimization with multiple starting points. We further propose an EM-based approach to estimate specimen-dependent measurement noise, tailoring the optimization to the specific specimen being imaged. The model predicts that optimized patterns achieve equal estimation accuracy in approximately 42\% fewer time steps compared to naive patterns.

Experiments on a real TEM set-up confirm the model predictions. The realized estimation variance across all tested tilt patterns closely matches the KF predictions, and optimized patterns substantially outperform naive ones. Beyond comparing tilt patterns, we also evaluate the overall method against the widely used Zemlin tableau. The proposed method achieves comparable or higher image quality on amorphous specimens. Importantly, the method also extends to non-amorphous specimens where the Zemlin tableau cannot operate, as demonstrated on a polycrystalline gold specimen.

A potential limitation is the reliance on the physical tilt-shift model, though the close match between predicted and realized estimation variance indicates limited model mismatch. The method also requires sufficient image features for reliable shift tracking, which may not hold for specimens with very uniform contrast. In such cases, the image tracker provides a natural diagnostic: anomalously small cross-correlation peaks between the template and the image can signal unreliable shift estimates that can be flagged or discarded. Beyond the TEM application, the sensor scheduling formulation and receding-horizon optimization approach may be of independent interest in other settings with continuous scheduling parameters and drifting system properties.

Future work could explore several promising directions. First, computational efficiency of the tilt sequence optimization could be improved by exploiting symmetries such as rotational invariance of the tilts or constraining the search space based on observed patterns in optimized tilt sequences. Second, extending the optimization problem to include more accurate modeling of the image shift tracker could lead to the removal of the ad-hoc tilt magnitude constraints included to allow the tracker to initialize. Finally, data-driven sensor scheduling approaches could learn high-performing tilt policies directly from experimental data, potentially bypassing the need for exact analytical models.


\bibliographystyle{elsarticle-num-names}
\bibliography{references.bib}

\end{document}